\documentclass[11pt]{article}

\usepackage[paper=a4paper,dvips,top=2.0cm,left=2.2cm,right=2.2cm,bottom=2.9cm,footskip=1.4cm,voffset=1.2cm]{geometry}
\usepackage{amsfonts,amsmath,amsthm,amssymb}
\numberwithin{equation}{section}  %%% Changes the equation numbering according to the section number
\usepackage[svgnames]{xcolor}
\usepackage{fix-cm}
\usepackage{sectsty}
\usepackage{fancyhdr}
\usepackage{lastpage}
\usepackage{graphicx}
\usepackage{url}
\usepackage{enumerate}
\usepackage{paralist}
\usepackage{multicol}
\usepackage{color}  %%% Needs to make colour of some content of your text
\usepackage{float}  %%% Fixes the position of the figure
\usepackage{epsfig}
\usepackage{hyperref}   %%% Needs to make hyper reference of any references or citations
\hypersetup{colorlinks=true,linkcolor=beamer@PRD, citecolor=beamer@PRD}
\usepackage{authblk}  %%% Helps to arrange author and institution details.
\usepackage{cite}  %%% Colour the citation brackets

\newcommand\myref[1]{\textcolor{beamer@PRD}{(}\ref{#1}\textcolor{beamer@PRD}{)}}

\definecolor{beamer@blue}{RGB}{0,0,255}
\definecolor{beamer@mediumblue}{RGB}{0,0,190}
\definecolor{beamer@midnightblue}{RGB}{25,25,112}
\definecolor{beamer@navy}{RGB}{0,0,128}
\definecolor{beamer@darkblue}{RGB}{0,0,139}
\definecolor{beamer@purple}{RGB}{128,0,128}
\definecolor{beamer@levander}{RGB}{100.,149.,237.}
\definecolor{beamer@PRD}{RGB}{46,48,146}
\definecolor{beamer@green}{RGB}{0,128,0}
\definecolor{beamer@darkgreen}{RGB}{0,100,0}
\definecolor{beamer@olive}{RGB}{128,128,0}
\definecolor{beamer@darkolivegreen}{RGB}{85,107,47}
\definecolor{beamer@gray}{RGB}{190,190,190}
\definecolor{beamer@ivry}{RGB}{220,220,220}%{238,232,205}
\definecolor{beamer@new}{RGB}{40,120,50}
\definecolor{shadecolor}{RGB}{220,220,220}
\definecolor{beamer@darkslategray}{RGB}{47,79,79}
\definecolor{beamer@chocolate}{RGB}{210,105,30}
\definecolor{beamer@brown}{RGB}{165,42,42}
\definecolor{beamer@orangered}{RGB}{255,69,0}
\definecolor{beamer@maroon}{RGB}{128,0,0}
\definecolor{beamer@white}{RGB}{234,242,243}
\definecolor{beamer@silver}{RGB}{0.5,0.45,0.37}

\begin{document}

%%%%%%%%%%%%%%%%%%%%%%%%%%%%%%%%%%%%%%%%%%%%%%%%%%%%%%%%%%%%%%%%%%%%%%%%%%%%%%%%%%%%%%%%%%%%%%%%%%%%%%%%%%%%%%%%%%%%%
%  Title
%%%%%%%%%%%%%%%%%%%%%%%%%%%%%%%%%%%%%%%%%%%%%%%%%%%%%%%%%%%%%%%%%%%%%%%%%%%%%%%%%%%%%%%%%%%%%%%%%%%%%%%%%%%%%%%%%%%%%%

\title{\textbf{On completeness of coherent states in noncommutative spaces with generalised uncertainty principle}}
\author{\textbf{Sanjib Dey} \\ \small{Centre de Recherches Math{\'e}matiques, Universit{\'e} de Montr{\'e}al \\ Montr{\'e}al H3C 3J7, Qu{\'e}bec, Canada \\ E-mail: dey@crm.umontreal.ca \\ \vspace{0.3cm} \textit{Dedicated to the memory of Prof. S. Twareque Ali}}}
\date{}
\maketitle
    	
%%%%%%%%%%%%%%%%%%%%%%%%%%%%%%%%%%%%%%%%%%%%%%%%%%%%%%%%%%%%%%%%%%%%%%%%%%%%%%%%%%%%%%%%%%%%%%%%%%%%%%%%%%%%%%%%%%%%%%
%  Abstract
%%%%%%%%%%%%%%%%%%%%%%%%%%%%%%%%%%%%%%%%%%%%%%%%%%%%%%%%%%%%%%%%%%%%%%%%%%%%%%%%%%%%%%%%%%%%%%%%%%%%%%%%%%%%%%%%%%%%%%%%
\thispagestyle{fancy}
\begin{abstract}
Coherent states are required to form a complete set of vectors in the Hilbert space by providing the resolution of identity. We study the completeness of coherent states for two different models in a noncommutative space associated with the generalised uncertainty relation by finding the resolution of unity with a positive definite weight function. The weight function, which is sometimes known as the Borel measure, is obtained through explicit analytic solutions of the Stieltjes and Hausdorff moment problem with the help of the standard techniques of inverse Mellin transform.
\end{abstract}	 
%%%%%%%%%%%%%%%%%%%%%%%%%%%%%%%%%%%%%%%%%%%%%%%%%%%%%%%%%%%%%%%%%%%%%%%%%%%%%%%%%%%%%%%%%%%%%%%%%%%%%%%%%%%%%%%%%%%%%%%
%  Introduction
%%%%%%%%%%%%%%%%%%%%%%%%%%%%%%%%%%%%%%%%%%%%%%%%%%%%%%%%%%%%%%%%%%%%%%%%%%%%%%%%%%%%%%%%%%%%%%%%%%%%%%%%%%%%%%%%%%%%%
\section{Introduction} \label{sec1}
\addtolength{\footskip}{-0.2cm} %% Adds extra space in the first page of the article
It is renowned that the coherent states are useful in different areas of modern science including quantum optics, atomic and molecular physics, mathematical physics, quantum gravity, quantum cosmology, etc, for further informations; see, for instance \cite{Ali_Antoine_Gazeau,Gazeau_Book}. Various generalizations of the Glauber coherent states have also become very popular in recent days giving rise to the possibility of constructing many new coherent states arising from various sophisticated mathematical backgrounds \cite{Perelomov,Nieto_Simmons,Filho_Vogel,Manko_Marmo_Sudarshan_Zaccaria,Gazeau_Klauder}. One of such prominent examples is the noncommutative space-time structure in the framework of generalised uncertainty principle, from which the existence of minimal length appears naturally \cite{Kempf_Mangano_Mann,Bagchi_Fring,Dey_Fring_Gouba,Dey_Fring_Khantoul}. There have been plenty of investigations behind the applications and usefulness of coherent states emerging out of the models on the noncommutative space \cite{Dey_Fring_squeezed,Dey_Fring_Gouba_Castro,Ghosh_Roy}. Furthermore, based on these coherent states, various nonclassical states have been constructed; such as, squeezed states \cite{Dey_Hussin}, Schr\"odinger cat states \cite{Dey,Dey_Fring_Hussin}, photon added coherent states \cite{Dey_Hussin_PhotonAdded} and their squeezing and entanglement properties have been studied. However, the mathematical completeness of such coherent states have not been studied before, which is important particularly to understand whether the coherent states are mathematically well-defined and can be utilized for the description of concrete physical systems.

The purpose of this article is to fill in this gap by finding the exact analytical expression for the positive definite Borel measure, such that the coherent states satisfy the required condition of resolution of identity. For this purpose, we mainly follow \cite{Klauder_Penson_Sixdeniers,Bergeron_Gazeau,Antoine_Gazeau_Monceau_Klauder_Penson} to associate our problem with the existing techniques of Stieltjes and Hausdorff moment problem and compute the inverse Mellin transforms corresponding to our systems, which yield the precise expressions of the Borel measure. The article is organised as follows: In Sec. \ref{sec2}, we introduce the basic notions of the generalised and nonlinear coherent states, as well as, the moment problem associated with them to identify the resolution of identity. In Sec. \ref{sec3}, we implement the existing framework as described in Sec. \ref{sec2} to study the completeness relation for the coherent states in noncommutative space for two different models, namely, the harmonic oscillator and the P\"oschl-Teller. Our conclusions are stated in Sec. \ref{sec4}.
%%%%%%%%%%%%%%%%%%%%%%%%%%%%%%%%%%%%%%%%%%%%%%%%%%%%%%%%%%%%%%%%%%%%%%%%%%%%%%%%%%%%%%%%%%%%%%%%%%%%%%%%%%%%%%%%%%%%%%
% Section 2
%%%%%%%%%%%%%%%%%%%%%%%%%%%%%%%%%%%%%%%%%%%%%%%%%%%%%%%%%%%%%%%%%%%%%%%%%%%%%%%%%%%%%%%%%%%%%%%%%%%%%%%%%%%%%%%%%%%%%
\section{Nonlinear coherent states and resolution of identity}\label{sec2}
We commence by revisiting the basic notions of nonlinear coherent states for the purpose of referencing. Nonlinear coherent states for Hamiltonians $H$ with discrete bounded below and nondegenerate eigenvalues, $E_n=\hbar\omega e_n=\hbar\omega nf^2(\hat{n})$, are defined as follows \cite{Filho_Vogel,Manko_Marmo_Sudarshan_Zaccaria,Sivakumar}
\begin{equation}\label{nonlinear}
\vert \alpha,f\rangle = \frac{1}{\sqrt{\mathcal{N}(\alpha,f)}}\displaystyle\sum_{n=0}^{\infty}\frac{\alpha^n}{\sqrt{\rho_n}}\vert n\rangle , \quad \rho_n=\displaystyle\prod_{k=1}^{\infty} e_k=n!f^2(\hat{n})!, \quad \rho_0=1,
\end{equation}
where $\alpha\in\mathbb{C}$ and, the normalisation constant can be computed from the requirement $\langle\alpha,f\vert\alpha,f\rangle =1$ as given by
\lhead{Completeness of coherent states in NC spaces with generalised uncertainty principle}
\chead{}
\rhead{}
\addtolength{\voffset}{-1.2cm} %% Adds extra space in the header of the first page of the article 
\addtolength{\footskip}{0.4cm} %% Adds extra space in the first page of the article 
\begin{eqnarray}\label{normalisation}
\mathcal{N}(\alpha,f) = \displaystyle\sum_{n=0}^{\infty}\frac{\vert\alpha\vert^{2n}}{\rho_n}.
\end{eqnarray}
In fact, nonlinear coherent states are the generalised versions of the Glauber coherent states \cite{Glauber} for the models corresponding to the generic function of the number operator $f(\hat{n})$. Note that, the term \textit{nonlinear} is not associated with the mathematical nonlinearity anyway, but it is because of its appearance in nonlinear optics. Mathematically the vectors $\vert\alpha,f\rangle$ in \myref{nonlinear} are well-defined in the domain $\mathcal{D}$ of allowed $\vert\alpha\vert^2$ for which the series \myref{normalisation} converges. The range of $\vert\alpha\vert^2,~0<\vert\alpha\vert^2<R$, is determined by the radius of convergence $R=\lim_{n\rightarrow\infty}\sqrt{\rho_n}$, which may be finite or infinite depending on the behaviour of $\rho_n$ for large $n$. Therefore, a family of such coherent states \myref{nonlinear} is an \textit{overcomplete} set of vectors in a Hilbert space $\mathcal{H}$, labelled by a continuous parameter $\alpha$ which belongs to a complex domain $\mathcal{D}$ (some domain in $\mathbb{C}$. For $R=\infty$, $\mathcal{D}=\mathbb{C}$). To be more precise, since $\vert n\rangle$ forms an orthonormal basis in the Hilbert space $\mathcal{H}$ and, let, $e_n$ be an infinite sequence of positive numbers, with $e_0=0$, then the vectors $\vert\alpha,f\rangle$ must satisfy the resolution of identity (completeness relation) with a weight function $\Omega$
\begin{eqnarray}\label{completeness}
\int\int_{\mathcal{D}} \frac{\mathcal{N}(\alpha,f)}{\pi}\vert\alpha,f\rangle\langle\alpha,f\vert~\Omega(\vert\alpha\vert^2)~d^2 \alpha = \mathbb{I}_\mathcal{H}.
\end{eqnarray}
By considering $\alpha=re^{i\theta}$, the left hand side of Eq. (\ref{completeness}) turns out to be
\begin{eqnarray}
&& \displaystyle\sum_{m,n=0}^\infty\frac{1}{2\pi\sqrt{\rho_m\rho_n}}\int_{0}^Rr^{m+n}\Omega(r^2)d(r^2)\int_{0}^{2\pi}e^{i\theta(m-n)}d\theta~\vert m\rangle\langle n\vert \\
&& ~~~~=\displaystyle\sum_{n=0}^\infty\frac{1}{\rho_n}\int_{0}^{R}t^n\Omega(t)dt~\vert n\rangle\langle n\vert,
\end{eqnarray}
such that one ends up with an infinite set of constraints
\begin{eqnarray}\label{measure}
\int_0^R t^{n}\Omega(t)dt = \rho_n, \qquad 0<R\leq \infty,
\end{eqnarray}
for which the completeness relation \myref{completeness} holds. Therefore, one can construct the coherent states \myref{nonlinear} for any models corresponding to a known $f(n)$, provided that there exists a measure $\Omega(t)$ which satisfies \myref{measure}. The explicit expression of the measure can be found, first, by associating \myref{measure} with the classical moment problem, where $\rho(n)>0$ are the power moments of the unknown function $\Omega(t)>0$ and, subsequently, by carrying out the integration by using the standard techniques of the Mellin transforms \cite{Oberhettinger}. For more details in this context, we refer the readers to \cite{Klauder_Penson_Sixdeniers,Antoine_Gazeau_Monceau_Klauder_Penson,Quesne,Ismail_Book}. For Glauber coherent states, i.e. for $f(n)=1$, $\rho_n=n!$, the moment problem \myref{measure} becomes 
\begin{eqnarray}\label{MeasureCanonical}
\int_0^\infty t^{n}\Omega(t)dt = n!, \quad n=0,1,2,.....,
\end{eqnarray}
so that one can easily identify the measure, $\Omega(t) = e^{-t}$. For $SU(1,1)$ discrete series coherent states \cite{Perelomov}, $\rho(n)=n!\Gamma{(2j)/\Gamma(2j+n)}$ and, the corresponding measure is given by
\begin{eqnarray}\label{MeasureSU11}
\Omega(t) = (2 j-1)(1-t)^{2j-2}, \qquad 
\end{eqnarray}
where $\Omega(t)$ is supported in the range $(0,1)$, with $j=1,1/2,2,3/2,3....$ In case of the Barut Girardello coherent states \cite{Barut_Girardello}, $\rho(n)=n!\Gamma(2j+n)/\Gamma(2j)$ and, the associated measure is given by the modified Bessel's function of second kind as follows
\begin{eqnarray}\label{MeasureBarut}
\Omega(t) = \frac{2}{\Gamma(2j)} t^{\frac{2j-1}{2}}K_{2j-1}(2\sqrt{t}),
\end{eqnarray}
where $\Omega(t)$ is supported in the interval $(0,\infty)$. For more examples of different types of coherent states; see, \cite{Klauder_Penson_Sixdeniers,Ali_Ismail}.  
%%%%%%%%%%%%%%%%%%%%%%%%%%%%%%%%%%%%%%%%%%%%%%%%%%%%%%%%%%%%%%%%%%%%%%%%%%%%%%%%%%%%%%%%%%%%%%%%%%%%%%%%%%%%%%%%%%%%%
%  Section 3
%%%%%%%%%%%%%%%%%%%%%%%%%%%%%%%%%%%%%%%%%%%%%%%%%%%%%%%%%%%%%%%%%%%%%%%%%%%%%%%%%%%%%%%%%%%%%%%%%%%%%%%%%%%%%%%%%%%%%
\section{Resolution of unity for coherent states in noncommutative space}\label{sec3}  
In this section, we will construct the coherent states arising from the noncommutative space \cite{Kempf_Mangano_Mann,Bagchi_Fring,Dey_Fring_Gouba}, in which the standard set of commutation relations for the canonical coordinates are replaced by the noncommutative versions, such as
\begin{equation}\label{NCOM}
[X,P]=i\hbar(1+\check{\tau}P^2), \quad X=(1+\check{\tau}p^2)x, \quad P=p,
\end{equation} 
where the noncommutative observables $X,P$ are represented in terms of the standard canonical variables $x,p$ satisfying $[x,p]=i\hbar$. $\check{\tau}=\tau/(m\omega\hbar)$ has the dimension of inverse squared momentum and $\tau$ is dimensionless. The above framework \myref{NCOM} is fascinating by itself, because it leads to the generalised version of the Heisenberg's uncertainty relation \cite{Kempf_Mangano_Mann} followed by the existence of minimal lengths \cite{Bagchi_Fring,Dey_Fring_Gouba}, which are one of the major findings of the string theory. Let us now discuss some concrete models in the given structure.

\subsection{Noncommutative harmonic oscillator}
We consider a one dimensional harmonic oscillator
\begin{equation}\label{NCHO}
H=\frac{P^2}{2m}+\frac{m\omega^2}{2}X^2-\hbar\omega\left(\frac{1}{2}+\frac{\tau}{4}\right),
\end{equation}
defined on the noncommutative space satisfying \myref{NCOM}. Here, the ground state energy is conventionally shifted to allow for a factorisation of the energy. The energy eigenvalues of \myref{NCHO} were computed in \cite{Kempf_Mangano_Mann,Dey_Fring_Gouba} by following the standard techniques of the Rayleigh-Schr\"odinger perturbation theory to the lowest order, as follows
\begin{equation}
E_n=\hbar\omega nf^2(n)=\hbar\omega n\left[1+\frac{\tau}{2}(1+n)\right]+\mathcal{O}(\tau^2).
\end{equation}
Correspondingly, by following \myref{nonlinear} the nonlinear coherent states are computed as
\begin{equation}\label{NCHOCS}
\vert \alpha,f\rangle_{\text{ncho}} = \frac{1}{\sqrt{\mathcal{N}(\alpha,f)}}\displaystyle\sum_{n=0}^{\infty}\frac{\alpha^n}{\sqrt{\rho_n}}\vert n\rangle , \quad \rho_n=n!f^2(n)!=\left(\frac{\tau}{2}\right)^n\frac{n!(n+\frac{2}{\tau}+1)!}{(1+\frac{2}{\tau})!}, \quad \rho_0=1.
\end{equation}
In order to verify that the states \myref{NCHOCS} are mathematical complete and well-defined in the Hilbert space, one needs to find out the existence of the positive definite Borel measure $\Omega(t)$ satisfying the constraint \myref{measure} as follows
\begin{eqnarray}\label{measureNCHO}
\int_0^\infty t^{n}\Omega(t)dt = \rho_n=\left(\frac{\tau}{2}\right)^n\frac{\Gamma(n+\alpha+1)\Gamma(n+\beta+1)}{\Gamma(1+\beta)}=\frac{A(n)B(n)}{\Gamma(1+\beta)},
\end{eqnarray}
with $A(n)=(\tau/2)^n, B(n)=\Gamma(n+\alpha+1)\Gamma(n+\beta+1)$ and $\alpha =0, \beta =1+2/\tau$. For the purpose of the computation of $\Omega(t)$ from \myref{measureNCHO} let us now briefly discuss some technicalities. First, we find the inverse Mellin transforms of the two functions $A(n)$ and $B(n)$ separately, i.e.
\begin{alignat}{1}
A(n) &=\left(\frac{\tau}{2}\right)^n=\int_0^\infty \left(\frac{t}{2}\right)^n\delta(t-\tau)dt=\int_0^\infty x^n C(x) dx, \\
B(n) &= \Gamma(n+\alpha+1)\Gamma(n+\beta+1)=\int_0^\infty x^n D(x) dx,
\end{alignat}
with 
\begin{alignat}{1}
C(x) &= 2 \delta (2x-\tau), \qquad D(x) = 2x^{\frac{\alpha+\beta}{2}}K_{\alpha-\beta}(2\sqrt{x}),
\end{alignat}
where $K_\alpha(x)$ denotes the modified Bessel function of second kind. Then, we utilize the composition formula \cite{Bergeron_Gazeau} to find the inverse Mellin transform for the composite system as given by
\begin{alignat}{1}
A(n)B(n) &=\int_0^\infty x^n \lambda(x) dx, \quad \lambda(x) = \int_0^\infty C(u) D\left(\frac{x}{u}\right) \frac{du}{u},
\end{alignat}
which when replaced in \myref{measureNCHO}, we obtain the accurate expression of the Borel measure $\Omega(t)$ as follows
\begin{eqnarray}
\Omega(t) &=& \frac{1}{\Gamma(1+\beta)} \int_0^\infty C(t)D\left(\frac{t}{u}\right)\frac{du}{u} \\
&=& \frac{1}{\Gamma(1+\beta)} \int_0^\infty \frac{4}{u}\left(\frac{t}{u}\right)^{\frac{\alpha+\beta}{2}} K_{\alpha -\beta}(2\sqrt{\frac{t}{u}})\delta(2u-\tau) du \\
&=& \frac{2^{\frac{1}{2}(4+\alpha+\beta)}}{\tau\Gamma(1+\beta)}\left(\frac{t}{\tau}\right)^{\frac{\alpha+\beta}{2}}K_{\alpha-\beta}(2\sqrt{\frac{2t}{\tau}}).
\end{eqnarray}

\subsection{A P\"oschl-Teller model in noncommutative space}
Let us now consider another interesting Hamiltonian based on the noncommutative space satisfying \myref{NCOM}
\begin{equation}\label{HamPT}
H_{\text{PT}} =\frac{\epsilon}{2m}P^2+\frac{\hbar\omega\gamma}{2\check{\tau}}P^{-2}+\frac{m\omega^2}{2}X^2+\frac{\hbar\omega\gamma}{2}+\frac{\epsilon}{2m\check{\tau}}, \qquad \gamma, \epsilon \in \mathbb{R}.
\end{equation}
Although the model \myref{HamPT} does not belong to a familiar class of models, however, it is very interesting as it leads to the well-known P\"oschl-Teller potential when the noncommutative observables are represented in terms of the standard canonical variables by using \myref{NCOM}. Therefore, one can describe the model as a noncommutative version of the P\"oschl-Teller model, although the Hamiltonian \myref{HamPT} can not be viewed as a deformation of a model on standard commutative space in the sense that it does not reduce to the usual P\"oschl-Teller model in the commutative limit $\tau\rightarrow 0$. For further details on the model; see \cite{Dey_Fring_Khantoul}, where the eigenvalues of the corresponding Hamiltonian were computed in an exact manner as given below
\begin{equation}
E_n = \frac{\hbar\omega\tau}{2}(1+2n+a+b)^2, \quad a=\frac{1}{2}\sqrt{1+\frac{4\gamma}{\tau}},~b=\frac{1}{2}\sqrt{1+\frac{4\epsilon}{\tau}}.
\end{equation} 
The corresponding nonlinear coherent states can be computed by following \myref{nonlinear} with 
\begin{equation}\label{RhoPT}
\rho_n = 2^n\tau^n\frac{\Gamma(n+\eta)^2}{\Gamma(\eta)^2}, \qquad \eta =\frac{3+a+b}{2}.
\end{equation}
Note that, the form of $\rho_n$ in \myref{RhoPT} is very similar to the case of harmonic oscillator as in \myref{NCHOCS} and, thus, we follow the similar procedure as discussed in the previous section to calculate the explicit expression of the measure
\begin{equation}
\Omega(t) = \frac{\tau^{-\eta}}{\Gamma(\eta)^2}\left(\frac{t}{2}\right)^{\eta -1}K_0(\sqrt{\frac{2t}{\tau}})~.
\end{equation}
%%%%%%%%%%%%%%%%%%%%%%%%%%%%%%%%%%%%%%%%%%%%%%%%%%%%%%%%%%%%%%%%%%%%%%%%%%%%%%%%%%%%%%%%%%%%%%%%%%%%%%%%%%%%%%%%%%%%%
%  Section 5
%%%%%%%%%%%%%%%%%%%%%%%%%%%%%%%%%%%%%%%%%%%%%%%%%%%%%%%%%%%%%%%%%%%%%%%%%%%%%%%%%%%%%%%%%%%%%%%%%%%%%%%%%%%%%%%%%%%%%
\section{Conclusions}\label{sec4} 
Coherent states in noncommutative spaces related to the generalised uncertainty relation have found to be interesting and useful for many different purposes \cite{Dey_Fring_squeezed,Dey_Fring_Gouba_Castro,Ghosh_Roy,Dey_Hussin,Dey,Dey_Fring_Hussin,Dey_Hussin_PhotonAdded}. However, it was necessary to find out their resolution of identity by computing the weight functions associated with them to show the mathematical completeness of the corresponding models. In this article we study the missing link by finding the completeness relations of coherent states for the harmonic oscillator and the P\"oschl-Teller model based on the noncommutative structure. We compute exact analytical expressions for the weight functions through the solutions of Stieltjes and Hausdorff moment problem for the two cases and show that the coherent states in noncommutative space are, indeed, mathematically well-defined and form a complete set of vectors in the Hilbert space. 

Evidently, there are many interesting open challenges left which will directly follow our results. By utilizing the Hankel determinant method, or any other existing mechanism in the literature \cite{Ismail_Book,Ali_Ismail,Englis_Ali,Dai} one may study the orthogonal polynomials, which are associated with our coherent states. The investigation may end up with some known orthogonal polynomials, or some fascinating $q$-orthogonal polynomials. However, because of the sophistication of our structure, the outcome may bring more exciting possibilities of constructing some new orthogonal polynomials out of our systems.  

\vspace{0.5cm} \noindent \textbf{\large{Acknowledgements:}} The author is supported by the Postdoctoral Fellowship by the Laboratory of Mathematical Physics of the Centre de Recherches Math{\'e}matiques.
%%%%%%%%%%%%%%%%%%%%%%%%%%%%%%%%%%%%%%%%%%%%%%%%%%%%%%%%%%%%%%%%%%%%%%%%%%%%%%%%%%%%%%%%%%%%%%%%%%%%%%%%%%%%%%%%%%%%%
%  References
%%%%%%%%%%%%%%%%%%%%%%%%%%%%%%%%%%%%%%%%%%%%%%%%%%%%%%%%%%%%%%%%%%%%%%%%%%%%%%%%%%%%%%%%%%%%%%%%%%%%%%%%%%%%%%%%%%%%%

%\bibliographystyle{unsrt} 
%\bibliography{Ref.bib}
%\input{Reference.tex}

%%%%%%%%%%%%%%%%%%%%%%%%%%%%%%%%%%%%%%%%%%%%%%%%%%%%%%%%%%%%%%%%%%%%%%%%%%%%%%%%%%%%%%%%%%%%%%%%%%%%%%%%%%%%%%%%%%%%%
%  Rulling
%%%%%%%%%%%%%%%%%%%%%%%%%%%%%%%%%%%%%%%%%%%%%%%%%%%%%%%%%%%%%%%%%%%%%%%%%%%%%%%%%%%%%%%%%%%%%%%%%%%%%%%%%%%%%%%%%%%%%
\vspace{0.3cm}
\begin{center}
\rule{2.5cm}{1.0pt}
\end{center}
%%%%%%%%%%%%%%%%%%%%%%%%%%%%%%%%%%%%%%%%%%%%%%%%%%%%%%%%%%%%%%%%%%%%%%%%%%%%%%%%%%%%%%%%%%%%%%%%%%%%%%%%%%%%%%%%%%%%%

\end{document}